\documentclass[%
 ,twocolumn%
 ,secnumarabic%
,amssymb, amsmath,nobibnotes, aps, prl]{revtex4}

\begin{document}

\title{On the problem of the correct interpretation
of phason elasticity
in quasicrystals}

\author{Gerrit Coddens}

\address{Laboratoire des Solides Irradi\'es,
Ecole Polytechnique,\\
F-91128-Palaiseau CEDEX, France}

\date{today}

\begin{abstract}
Recently Francoual {\em et al.} \cite{Fracoual}
claimed to have observed the dynamics
of long-wavelength phason fluctuations in
{\em i}-AlPdMn quasicrystals.
We will show that
the data reported
call for a more detailed development of
the elasticity theory
of Jari\'c and Nelsson\cite{Jaric}
in order to determine the
nature of small phonon-like atomic displacements
with a symmetry that follows the phason
elastic constants.
At the end of our paper we present the
reader with a discussion,
where we rebut some recent objections.
\end{abstract}


\maketitle

\narrowtext

Recently Francoual {\em et al.} \cite{Fracoual}
claimed to have observed the dynamics
of long-wavelength phason fluctuations in
{\em i}-AlPdMn quasicrystals.
These claims were based on the observation
of very  long
relaxation times in the so-called speckle
patterns
of the diffuse scattering measured
with coherent X-rays.

These claims run contrary to the fact that
the diffuse scattering and/or its kinetics
cannot possibly correspond to the
flipped tile configurations that occur in the
Monte Carlo random tiling
simulations.\cite{Tang,Shaw}
In fact, in this
model the number of flipped tiles,
and correspondingly the diffuse scattering intensity,
should increase with
temperature, while the experimental data
exhibit precisely the
opposite temperature
dependence. The authors are aware
of this: In
reference\cite{ISIS} they stated
that the data are ``in contradiction
with the hypothesis
of a simple random tiling model''.
Instead of considering
the very obvious possibility
that this indicates
that the data have nothing to do with tile flips,
the authors
added that the tile flip interpretation
can be maintained by
introducing a more complex random
tiling model.\cite{Widom}
The random tiling model  and
the tile flip scenario are thus maintained
at the prize of swapping models with completely
opposite predictions about the temperature behaviour.
This creates the impression that the
interpretation of the data was
preconceived.
To cite Sir Karl Popper:
{\em ``Falsifiability should be the criterion
  of demarcation in science''}.\cite{Popper}

The diffuse scattering data, without a mention
of their temperature
dependence, were eventually reported
  in reference \cite{wrong1}.
Nonetheless, in the final discussion of that paper,
the authors already introduce the arguments to
thwart the criticism that can be formulated based
on the temperature behaviour  observed. As the reader
had no access to the information about the  temperature
dependence, he could not possibly understand
the issues at stake.
Rather than the mention of a
crucial problem raising serious doubts\cite{IJMPB}
about the validity of the interpretation,
it looked like a very puzzling
digression. Moreover,
all the temperature data might show is
that there is some softening of the elastic constants.
There are many examples
known of elastic constants or phonon  modes
that  soften in a given
temperature range,
without triggering any phase transition
at all.
In their paper, the authors concluded
that their data were ``compatible'' with the
random tiling
model. The problem with this statement does
not reside in what it
literally states, but in the  crucial piece
of information it tacitly omits,
viz. that the data may also be compatible
with other models.
Due to this formulation the reader may pick
  up the totally
unjustified belief that the data would have
conclusively proved the random tiling model.
As a matter of fact, such overinterpretations have indeed
made their way to the literature,\cite{Elser}
installing a strong oral tradition, which has
never been properly eradicated.
The unassailable character
of the literal interpretation of the formulation
may have contributed to this situation, by
facing the reader with
a {\em de facto} denial of the possibility to take issue.
This longstanding absence of a published rebuttal
  cannot mask the real situation: The data
do not contain any evidence
that would warrant such claims.
And it is not by qualifying this problem
as an old story,
\cite{private} that the validity of these claims
can be settled.

Because the data actually {\em contradict}
an interpretation
in terms
of a signal corresponding to structural
disorder obtained by tile flips,
as the temperature dependence of the data
already mentioned clearly indicates.
The authors explicitly announce their
awarenes of the fact that
the word
phason is ambiguous as it has been used with
several different meanings,
while it is the transgression of this very
caveat that tacitly
serves as a platform for their unjustified claims,
by confusing two different meanings of
the terminology phason:
The existence of
a terminology ``{\em phason} elasticity'' and
the fact that their model can be called
a `` {\em random tiling} model''
are used to suggest without any proof
that what they observe would be {\em phason} flips.
Rather than being able to prove the random tiling
scenario the authors had great difficulties
in saving the random tiling paradigm and were forced to
introduce several {\em ad hoc} assumptions in order
  to achieve this.
This rescue operation works only in the appearances.
In fact, as we already indicated, the intensity
of the quasielastic neutron scattering signal that
corresponds to the tile flips increases when
the temperature is raised. This is model-independent
factual information. If the
diffuse scattering observed by the authors
were to correspond to tile flip kinetics
it should follow the same temperature behaviour
as in the neutron data
rather than the opposite one.

Perhaps, this requires a more detailed discussion.
If one were able to diagonalize
the huge jump matrix that describes the whole of the phason
dynamics one would find a (very large) number
of characteristic times,
each leading to a Lorentzian signal with a dynamical
structure factor. Such a jump matrix exists both
for the coherent and for the incoherent scattering case.
Such a theoretical treatment is beyond reach however,
due to the sheer size and complexity of configuration space.
The observed neutron scattering data correspond to the sum
of coherent and incoherent scattering (contrary to
the statements of
Fracoual et al. that the neutron scattering
signals would be incoherent).
  The long-time signals attributed to
phason dynamics by
Fracoual et al. would just correspond to
some of these Lorentzians,
with very long relaxation times.
These long relaxation times are not elementary
parameters of the jump model, but functions
of more elementary
jump times. These functions
pop up as the inverses of the eigenvalues
$\lambda_{j} = f_{j}(\tau_{1}, \cdots \tau_{n})$
of the jump matrix
that has been defined in terms of the few more
elementary jump times, $\tau_{1}, \cdots \tau_{n}$,
which are much faster. The
Q-dependence of the intensities
of all Lorentzians $\Lambda_{j}$ is given by
the corresponding structure factors.
Any temperature dependence enters into the model
through the temperature dependence
of the elementary jump times in terms of
activation energies.
Through the functional dependence
$\lambda_{j} = f_{j}(\tau_{1}, \cdots \tau_{n})$ evoked,
the temperature dependence of all Lorentzians
is thus dictated by the temperature dependence
of the elementary jump times.
According to crude criteria such as
increasing or decreasing
of jump times or intensities, the long time dynamics
should thus have the same temperature dependence
as the short time dynamics, and any person who wants
to formulate a claim that they could show
opposite behaviour
will have to work very hard to gain
credibility for it.

But the experimental observation
(from quasi-elastic neutron scattering)
that the intensity,
rather than the width
of the fast signals increases with temperature
is unusual.
In a first approach, one might argue that
it could be due to the finite energy resolution
of the neutron scattering experiments:
At low temperatures
the dynamics are too slow to be resolved and appear
as elastic. At higher temperatures they
become resolved
leading to the illusion that the intensity
of the elastic peak
decreases and the intensity of the
quasielastic signals
increases accordingly. But this is not what
happens:
The width of the quasielastic signals
cannot be detected
to change with temperature,
while their intensities change drastically
in a way that
cannot be attributed to some broadening.
One needs to introduce an assistance scanario
to explain this
very unusual behaviour.

But what Fracoual et al. observe
and call the effect of an inverse Debye-Waller factor,
is that the intensities (i.e. structure factors)
of the very slow signals {\em decrease} while
the temperature is raised,
{\em with the intensities being transfered to
the Bragg peaks!}
If there were no assistance scenario in the jump dynamics,
the intensity and its Q-dependences for a given
Lorentzian $\Lambda_{j}$ would remain
the same at all temperatures.
In fact, the structure factor of the
Lorentzian $\Lambda_{j}$
associated with
$\lambda_{j}$ is not changed by
a speeding up of the dynamics
(allowance made for the phonon
Debye-waller factor).
At the very best, its intensity at a given Q-value
would appear to be associated with a
{\em faster} relaxation time,
through the functional relationship
$\lambda_{j} = f_{j}(\tau_{1}, \cdots \tau_{n})$.
If one wants to change $f_{j}$ or the
structure factors, rather than just
$\tau_{1}, \cdots \tau_{n}$, one has to
introduce special assumptions.
The observation that the quasielastic
neutron scattering intensity
increases with temperature forced us to
introduce such an assumption in the
form of an assistance
scenario, as even the elastic
intensity is determined by
the jump model: It corresponds to the
eigenvalue $0$ of the
jump matrix, and hence its intensity or
structure factor should normally
not change with temperature. In other words:
Allowing for the effect of the Debye-Waller
factor due to the phonons,
the ratios of the
various structure factors, including the elastic one,
should have remained the same.
In the assistance scenario, the elastic
intensity decreases
with temperature, because one introduces
long-lived excited
states, whose population is governed by
a Boltzmann factor with a large
activation energy.
What kind of most extraordinary
{\em ad hoc} assumptions
would have to be introduced into the
jump model in order to
obtain a {\em decreasing} diffuse
intensity as observed by the authors,
that could be attributed to phason jumps
despite the fact that the
quasielastic intensity
corresponding to fast phason
jumps has been observed
to {\em increase} by neutron
scattering?

To resume the situation: The number of tiles that
flip increases when the temperature is raised.
Therefore, the intensity of the signal that
should betray the presence
of these tile flips, e.g. off-Bragg-peak
diffuse scattering
claimed to correspond to the structural
disorder produced by the flips,
should also increase when the temperature
is raised.\cite{remark}
Such a temperature dependence runs
contrary to what the authors observe.
They then decide to proceed by
postulating an ``alternative
random tiling model'', wherein
the diffuse scattering intensity decreases
when the temperature is raised.
Such an expedient does not change a iota to the
fact that the diffuse scattering
intensity cannot be attributed to
structural disorder
produced by tile flips as it has
the wrong temperature
behaviour. It follows that  the diffuse scattering,
which they call the ``phason fluctuations''
of the alternative random tiling model,
must be dissociated from tiling disorder,
i.e. the ``phason fluctuations'' are
not tile flip kinetics.

But the authors just ignore
this fact and introduce a new
terminology ``phason fluctuations'' to suggest
that  the interpretation
of the diffuse scattering
in terms of tile flip kinetics
would be saved by such an {\em ad hoc}
swapping of models.
This introduces several confusions.
(1) Perhaps the random tiling philosophy can be saved
by such a swapping, but not the interpretation
of the ``phason fluctuations'' in terms of tile flip
kinetics. (2) The introduction of the
new terminology ``phason fluctuations''
is also confusing. It works like
a  tacit  to and fro
swapping between two states of clarity
about the validity of what is being claimed.
As the temperature dependence shows,
an interpretation of the data in terms of
tile flip kinetics
is clearly wrong.
When the new terminology ``phason fluctuations''
is introduced,
its exact meaning is not well defined, such that
it is no longer clear
what kind of interpretation for the
data is being claimed.
Our verdict about the validity of
the interpretation
is then in suspense:
Perhaps the new terminology signals
that the  original scenario
in terms of tile flip kinetics is
being abandoned.
When the presence of the
word ``phason'' in the
terminology is used to reintroduce
the interpretation in terms of
tile flip kinetics, we swap back to
the context of the explicit original claims.
Those remain wrong but
the meandering path taken by the authors tends to
leave the reader
  with the (false)
impression that he  no longer
is able to be absolutely sure
that the interpretation
is  in disagreement
with the data.
The inverse Debye-Waller effect can
certainly not be just a matter
of tile flips as the authors claim.
The temperature dependence
clearly shows that some other phenomenon must be
responsible for the data,
something that has nothing
to do with the tile flip dynamics.

This is a towering microscopic-level
objection against
the phason interpretation of the
speckle signals.
But the authors do not address these
microscopic issues.
By formulating their claims in a
macroscopic language
of phason elasticity, they dress a
language barrier:
Nobody understands
how the macroscopic phason elasticity
is supposed to relate to the
microscopic-level tile flips, if it does at all.
And therefore, nobody knows how
the previously mentioned
microscopic objections
should be translated across this
language barrier.
Moreover, this phason elasticity
scenario is presented
as though it would be conceptually
clear and self-evident
while it is not. In Widom's paper
it is stated that on lowering the temperature
the QC moves away from the ideal
random tiling conditions
and that this drives an elastic instability.
It is not told how this should be described
on the microscopic level. The elastic instability
could e.g. correspond to a distortion
of the tiles rather than to their mere flips.
Widom's paper talks about an inverse
Debye-Waller effect
on the {\em elastic} intensity.
Why has this to apply
for the diffuse scattering and not
for the Bragg peaks,
if it is true that the experimental
data confirm this?

To save the interpretation of the data
in terms of tile flips,
the authors try to fall back onto
critical fluctuations,
whose microscopic description is again eluded.
They do not address the question
if after their 180 degrees turn in the
interpretation
of the data, phason elasticity can
still be identified
in a 1-1-way with a macroscopic description
of  tile flips with no other ingredient at all,
like it appeared in the simple random
tiling interpretation
that was based
on Monte Carlo simulation of purely
entropic tilings.
There is no elaboration of ideas or proofs
in this at all.
The authors just swap models in a tacitly
assumed {\em tertium non datur},
while there is a very
obvious alternative that would be
worth investigating,
viz. that the diffuse scattering
intensity is not produced
by tile flips.

The loophole of escape proposed by the
authors cannot be correct.
First of all, there is no evidence
for the approach of a phase transition.
In fact, one must introduce
{\em ad hoc} assumptions
in order to deny the clear experimental evidence
that there is no phase transition.
And secondly, we can state that in general,
there is no guarantee for the attribution
of a wavelength to the quantity
$q$ as the authors do.
As there is no conclusive interpretation
of the diffuse scattering, considerations
about the nature of the {\em ad hoc} stipulated
purely hypothetical transition
(order-disorder, first or second order)
cannot play a role at this stage.
Consider thus the clear analogon
of a (second-order) antiferromagnetic
phase transition.
(Note that the authors cite themselves
the second-order phase transition
in NaNO$_{2}$ as an illustration of their views,
in contradiction with their own claims that
the transition should be first order).
At approaching
the N\'eel temperature from above, larger and larger
antiferromagnetically ordered
domains (or clusters) will occur that
will take longer and longer
times $\tau$ to decay. This will show
up as diffuse scattering
intensity centered at the antiferromagnetic
Bragg position, e.g. at
${\mathbf{Q}} = [{1\over{2}},0,0] {2\pi\over{a}}$ of
the future low-temperature phase,
where $a$ is the lattice parameter
of the high-temperature phase.
The intensity at ${\mathbf{Q}} + {\mathbf{q}}$ will
have a characteristic decay time $\tau$, which
is a measure of how long
an antiferromagnetically ordered cluster of
size $2\pi/q$ will persist in time
without being disrupted by the spin flip dynamics.
We see that it is $2\pi/Q$ rather than $2\pi/q$
that characterizes the wavelength $2a$
of the spin wave that is being built up. The quantity
$2\pi/q$ is not a long wavelength of some spin wave,
but an instantaneous domain size, a coherence length
of the short wavelength spin wave.
The time $\tau$ is
not characteristic of the spin flips
themselves (which are local),
but of the absence of spin flips within
a domain of size
$2\pi/q$. These domain sizes increase
when the spin flip
dynamics slow down on approaching $T_{N}$.

In this discussion, we use the
phase transition
only to illustrate a possibility
of an interpretation.
In the context outlined above,
this possibility will remain valid
in its general ideas,
even if there is no phase transition at stake at all:
$q$ refers to a domain size, rather than
to a wavelength.
By analogy, we see that it is wrong in
the quasicrystal,
to associate $2\pi/q$ with some
hypothetical long wavelength
phason wave. Moreover, if there were
some wavelength $\lambda$
in the phason dynamics,  diffuse
scattering should build maxima
at {\em new} Bragg peak
positions at ${\mathbf{Q}}$ with
$Q=2\pi/\lambda$,
or at satellites positions at
${\mathbf{Q}}= {\mathbf{G +q}}$
rather than remaining smeared out over
a continuum of positions ${\mathbf{G +q}}$
  in a distribution centered on
${\mathbf{G}}$, with
$q=2\pi/\lambda$ (if it is the signature
of the mechanism behind the transition).
As the diffuse scattering in QCs remains centered at
the Bragg peaks of the high temperature regime
and its maxima do not define new
Bragg or satellite positions,
the {\em ad hoc} interpretation in
terms of critical scattering
announcing a phase transition that
would not be
reached due to the slowing
down of the phason dynamics, is wrong.
Note that the slowing down of the spin flips
is what triggers the antiferromagnetic
phase transition
rather than impeding it!

The battle horse of the authors to escape
from these objections might well
be the superstructure reported by
Ishimasa,\cite{Ishimasa} which shows both
sattelite peaks and diffuse scattering.
However, for this
exceptional observation there are many
other ones where
there is no phase transition at all.\cite{wrong2}
But even in the favorable case reported by
Ishimasa, the dominant
diffuse scattering maxima are centered
on the Bragg positions of the QC, not on
the sattelites.
The scenario is thus not one of diffuse scattering
progressively building up at the future
satellite positions, and
eventually turning into sattelite Bragg peaks.

Now that we have shown that  both
(quite opposite) random tiling
scenarios \cite{Tang,Widom}
do not agree with the data, we are left
with the problem
of providing their correct interpretation.
This is not a logical necessity
for establishing that the interpretation
of the authors is wrong, but we fear
that it might be argued that
alternative interpretations that are not beset with
similar errors are not at hand. In fact,
the bold claims
of the authors have installed a
reversal of the charge of proof.
The wrong interpretation and  claims
will keep hanging around until
someone will
have definitively solved the
very hard problem of the correct interpretation
of the diffuse scattering,
and provided watertight proof for it,
making cautiously allowance
for all possible objections.
While such a final, unambiguous solution is
far beyond our possibilities
we will nevertheless sketch a few arguments
to convince the reader
that an alternative interpretation is
not at all impossible.
We will argue below, that exactly the observation
of the kinetics further excludes the possibility
of an interpretation of the phason elasticity
in terms of tile flips, as atomic jumps cannot
be correlated over large distances, as the authors
assume. The arcane character of this
counter-intuitive assumption
should at least have been spelled out
with great emphasis by the authors rather
than presenting it as though
  it were a trivial possibility, that does not
  warrant any discussion or justification.
This is a most obvious objection,
which only reflects standard common sense in
atomic self-diffusion studies, but
de Boissieu just waves it aside by qualifying it
as pure speculation.\cite{speculation}

At this point it is perhaps good to
review the evidence
the authors proposed in favor of their
interpretation
in terms of tile flips. (a) An $1/q^{2}$
dependence
of the intensity. This is unspecific
and occurs in many
other cases of Huang scattering in
crystals.\cite{Salje}
(b) A certain shape
of the diffuse scattering intensity contours.
Again, this is unspecific, and purely
due to symmetry
in analogy with the situation in
conventional crystals.\cite{Salje}
Such symmetry-based arguments do not
contain any information
about the underlying mechanisms or
interactions. (c)
The shapes of the contours depend only
on the phason-phason
elastic  constants.
This is the only feature that is not
unspecific. As such,
it may rule out a number of alternative
models.\cite{warning}
But this does not mean  that
it rules out all possible alternative
models, and that an
interpretation in terms of random tiling
configurations would be unique, even
if the authors
try to present this idea as self-evident.
Such a way of presenting the state of affairs
is in flagrant contradiction with
the attitude that might transpire from the
{\em caveats} the authors issue towards
the reader that the word
``phason'' has been used with several
different meanings.
In fact, it is not because
both the elastic constants and tile
flip kinetics
can be labeled with the word ``phason''
that
we would have a direct relationship
between the two phenomena.
As the temperature dependence does not agree
with the random tiling scenarios, we
actually know that
the correct model must be different.
We will point out that other phenomena
than tile flips can lead to
``phason elasticity'', such that there
is not a 1-1-correspondence.

We should
insist on the fact that an interpretation
in terms of tile flips
is a {\em derived} application of the
elasticity theory,
which is formulated in terms of a
continuum
of small atomic displacements rather
than on a discrete set
devoid of infinitesimals. The very
definition
of an elastic constant cannot be written down
if we cannot assume that the atomic
displacements
explore a continuum. The validity of the derived
application is not
obvious, as on the microscopic level
tile flips do not explore a continuum
of atomic displacements.
There is thus {\em a priori} no good
theoretical rationale to explain the results
of Tang except the {\em post facto} observation
that it works despite such theoretical objections.
To improve on this situation,
Henley\cite{Henley} has proposed an argument
in terms of coarse graining,
but we find this rather vague
and would prefer a more precise
mathematical description. As a
corollary, we think
that it would be conceptually much more clear
if even on the microscopic
level we could have a continuum of small
atomic displacements. In any case, it remains
  a cracking pass to generalize
the finding by Tang in the sense that one
takes it for granted that the only exclusive way
to obtain such a dependence on the phason
elastic constants would be the tile flips
that are so improper for the
first-hand application
of the theory. Similarly, in Widom's
Landau-type theory
it is not granted that the phason elastic constants
can only correspond to tile flips:
Even if some elastic
instability were observed in diffraction
experiments that
completely tallied with
his calculations, it would not yet
prove the random tiling scenario.

The situation with a physical theory is somewhat
analogous to the one in analytical geometry.
The geometrical information is coded
by a 1-1-mapping
into an algebraic formulation and the
theorems are obtained by making the
calculus and translating
the final results backwards into geometry.
Going backwards and forwards between the algebra
and the geometry all along the development
can be a very
beautiful and revealing experience.
In physics, we have a similar coding of the
physical phenomena
in the form of a calculus.
The problem with the theory of Jari\'c and
Nelsson\cite{Jaric}
is that it provides only the calculus and not
the coding. It leaves the problem of cracking
the code with the reader. Making good
physical sense
of the algebra is difficult, since
the elasticity theory is a continuum theory,
while the quasicrystal is a discrete set of
atomic positions.
What the microscopic interpretation of the continuum
theory should be has never been very much debated,
let alone that it would have been conclusively
settled. Nonetheless, the authors tacitly introduce
a very specific microscopic interpretation
without any proof, and move on further as
though it were
self-evident, well established knowledge.

One way, admittedly not the most clever
one, to crack a code
is guessing. It is then necessary to check if
the guess makes proper sense. Promoting
the guess to dogma without checking will
most of the time
not lead anywhere.
The authors' guess is that in a
simplified picture,
phason waves can be viewed
as sine waves propagating in the physical
space with a polarization
in the perpendicular space. That guess
sounds conceptually familiar and clear,
but on checking if it makes sense it
looks badly wrong.
A physical wave is generally of the form
${\mathbf{u}}({\mathbf{r}}) =
{\mathbf{u}}_{0} \sin({\mathbf{q\cdot r}}+
\varphi)$.
In QCs this could be
${\mathbf{u}}_{\parallel}({\mathbf{r}}) =
{\mathbf{u}}_{0 \parallel}
\sin({\mathbf{q\cdot r}}+ \varphi)$.
What the authors propose is rather:
${\mathbf{u}}_{\perp}({\mathbf{r}}_{\parallel})
= {\mathbf{u}}_{0 \perp}
\sin({\mathbf{q\cdot r}}_{\parallel}+ \varphi)$.
This leads to
${\mathbf{u}}_{\parallel}({\mathbf{r}}_{\parallel}) =
{\cal{G}}(u_{0\perp}
\sin({\mathbf{q\cdot r}}_{\parallel}+ \varphi))$,
where ${\cal{G}}$
is a function that translates the sine wave in
terms of parallel-space
displacements. The function ${\cal{G}}$ is
not analytical. Infinitesimal
displacements in perp-space lead either
to zero displacements
or to discrete finite
phason jumps. In other words ${\cal{G}}$
cannot be approximated by
a linear mapping, such as to recover something like
${\mathbf{u}}_{\parallel}
({\mathbf{r}}_{\parallel}) = C_{1} + C_{2}
\sin({\mathbf{q\cdot r}}_{\parallel}+ \varphi))$,
with $C_{2} = d{\cal{G}}/dr_{\perp}$,
and perhaps $C_{1}= 0$.
Hence the sine wave proposed is not a wave.

This has many manifestations. One can choose
a sine wave of small
amplitude that will just define a few
isolated tile flips
with long distances in between.
The  atomic displacements involved
will explore only two values (viz. zero
and the jump distance)
instead of a continuum of values.
The effect of doubling the small amplitude
of the sine wave will not lead to a doubling
of the corresponding
displacement amplitudes in physical space.
And contrary to one's
first possible intuitions,
the displacement pattern defined
by the periodic sine wave is itself not periodic,
but aperiodic, just like the cut
defining the quasicrystal does not define
a periodic structure.
For large amplitudes the sine wave may intersect
the same atomic surface
more than once, which is also unphysical.
The sine wave will lead uniquely  to atomic jumps,
but an atomic jump never explores the harmonic regime
of the  potential, while  elasticity
theory is set up within the harmonic approximation.
And finally, long distance wave-like
correlations between atomic jumps
without any external driving force are
an awkward notion:
Atomic jumps are basically stochastic and
anharmonic
in nature, and that
a few jumps could perhaps be correlated
{\em locally and simultaneously}
was already quite surprizing.\cite{myPRL2}
A correlation between two jumps that are separated
by a long distance and a long time interval
just does not make physical sense
(especially when no atomic displacements
occur in between
as in the picture of a wave with
small amplitude
and a long wavelength).
A ``diffusive phason wave''
is a {\em contradictio in terminis}.

The ``simplified'' picture evoking a
phason wave,
is thus not at all as self-evident as suggested,
and the misleading evocation of a
pictorial simplicity helps
the reader in accepting
  the introduction of an ill-defined, flawed
  concept rather than
  making sense of a real physical phenomenon.
There exists no microscopic drawing in the literature
showing the postulated
simple picture of a phason wave in a
quasicrystal. There
also exists no
schematic pictorial description of the
diffusion kinetics of such a wave.
There exists even less a rationale
that would show how this diffusion
would lead to the signals that are attributed
to them.
All this is not because these matters would be
too trivial and self-evident to spend one's
time on explaining them,
as one might infer from the tacit style of presentation.
On the contrary, one runs into great difficulties when
one tries to justify or make sense of the
``simple picture''.
The whole is just postulated,
thereby transfering the charges of proof
(that it cannot be true) and of creative
thinking to those who wonder.
(We will discuss a related problem below,
viz. that reporting $\tau$ for
one selected set of $({\mathbf{Q}},{\mathbf{q}})$-values
has little physical meaning, and certainly does not
correspond to the ``simplied'' picture of ``phason
wave'' postulated by the authors.
It can only be a Fourier
component).

For phonons on a quadratic two-dimensional periodic
lattice we can imagine four types of distinct
phonon modes: Both the polarization and
the propagation vector can be along the $x$-
and the $y$-direction. One may guess that the analogon of this
for the quasicrystal leads to four basic
possibilities of propagation:
$u_{\parallel 0}\sin({\mathbf{q}}_{\parallel}
{\mathbf{\cdot r}}_{\parallel})$,
$u_{\parallel 0}\sin({\mathbf{q}}_{\perp}
{\mathbf{\cdot r}}_{\perp})$,
$u_{\perp 0}\sin({\mathbf{q}}_{\parallel}
{\mathbf{\cdot r}}_{\parallel})$,
$u_{\perp 0}\sin({\mathbf{q}}_{\perp}
{\mathbf{\cdot r}}_{\perp})$.
But we must make a distinction between
auxiliary and real physical
quantities.
At least in a first approach, the latter
two  possibilities
can be disconsidered,
as in general,  a small displacement along
perpendicular space does
not lead to any real-space atomic displacements,
except for a few isolated atomic jumps.
These phason jumps are the exception rather
than the rule
with respect to the basic concept that a
perpendicular space displacement
should not lead to a visible displacement in real space.
It would make thus much better sense  that a phason
wave would be of the second type, rather than of the
third type as the authors claim, but
we have to draw in still another consideration.

If this analogy with the phonons on a quadratic
lattice were strict,
$({\mathbf{r}}_{\parallel},{\mathbf{r}}_{\perp})$
would explore the set of nodes of the
superspace hypercubic lattice.
But QCs have the particularity that their
atomic surfaces
are not points but extended sets. This complicates
the simple analogy with the square lattice
developed above, since
there can now also be a dependence of
the amplitude of the displacements
on ${\mathbf{r}}_{\perp}$ along the atomic surface,
rather than just the
${\mathbf{r}}_{\perp}$-coordinate of the node.
Hence it is also possible to have e.g.
$u_{\parallel}({\mathbf{r}}_{\perp})
= u_{\parallel 0} f({\mathbf{r}}_{\perp})$,
where $f$ is a function on the atomic surface.
In the continuum theory, all such detailed
microscopic considerations
are lost as one reasons on average atoms.
The possibility
$u_{\parallel}({\mathbf{r}}_{\perp})
= u_{\parallel 0} f({\mathbf{r}}_{\perp})$
leads to real atomic displacements
whose amplitudes can explore a continuum,
rather than a discrete set,
such that a linearization of the theory for small
atomic displacements in the harmonic regime is feasible,
even without going from the microscopic to the
continuum regime.
In conclusion,  phason modes can be
fields of small parallel-space atomic displacements
whose amplitude is a {\em function} of the
perp-space coordinate
of the atomic position rather than its parallel-space
coordinate as is the case for a phonon,
e.g. $u_{\parallel}({\mathbf{r}}_{\perp})
= u_{\parallel 0} f({\mathbf{r}}_{\perp})$,
where $f$ is an analytical function on the atomic surface,
a linear one in the first order approximation.

In the further elaboration, this leads to
further possibilities,
e.g. $u_{\parallel}({\mathbf{r}}_{\parallel},
{\mathbf{r}}_{\perp})
= u_{\parallel 0} f({\mathbf{r}}_{\perp})
\sin({\mathbf{q}}_{\parallel}
{\mathbf{\cdot r}}_{\parallel})$.
The latter is a wave propagating along parallel space,
whose polarisation is a function of the perpendicular
coordinates. The polarisation is of course along parallel
space, as it should be. In the set of further
possibilities,
we now can recover the two possibilities
$u_{\perp 0}\sin({\mathbf{q}}_{\parallel}
{\mathbf{\cdot r}}_{\parallel})$,
$u_{\perp 0}\sin({\mathbf{q}}_{\perp}
{\mathbf{\cdot r}}_{\perp})$,
which we {\em a priori} had rejected,
as they now lead in general to real-space
displacements,
  since the introduction of $f$ has introduced a
  1-1-correspondence between
the perpendicular and the parallel coordinates.
Such a reintroduction of these two possibilities
is however, a mere reformulation of the two
more fundamental possibilities
$u_{\parallel 0}\sin({\mathbf{q}}_{\parallel}
{\mathbf{\cdot r}}_{\parallel})$,
$u_{\parallel 0}\sin({\mathbf{q}}_{\perp}
{\mathbf{\cdot r}}_{\perp})$,
through the bias of the 1-1-correspondence
introduced by $f$.
But due to this 1-1-correspondence,
the word polarization can  take now two
quite different meanings:
A derived one that refers to an undulation
of the cut as suggested by the authors,
and a more fundamental one that refers to
an undulation
of the atomic surfaces.\cite{note}
In the example of the Fibonacci chain,
we could e.g. tilt all atomic surfaces over a
small angle $\alpha$.
Such a displacement field
  would still lead to a diffraction pattern with
  the same Bragg peak positions.
Only the intensities would be changed.
If we imagine a fluctuating cut, we get a model
that is very similar to the one that has
been proposed by the authors,
but that now contains small atomic displacements, and
does not need to imply tile flips.
 From Tang's random tiling simulations,
we know that the cut
can be a totally random function rather
than a sine wave,
and still lead to diffuse scattering
  at ${\mathbf{Q+q}}$, while $q$ has nothing
  to do with a wavelength.
If the assumptions of the authors are able
recover all experimental
characteristics (a)-(c) outlined above, then
the same must be true
for our model of small atomic displacements.

To be quite exact, we do  not have to claim that
our model is correct. But we do have proved
that our model
definitely excludes the model proposed by the authors
from the list of possibilities.
The essential point is that the real-space
atomic diplacements created
by the fluctuation of the cut are much larger
and not harmonic in the model of the authors.
We think that our alternative leads to a less unphysical
interpretation of the data, without a real necessity to
invoke tile flips  as the basic ingredient.
The displacement fields are just like classical
phonon displacement fields, except for the fact that
they are parameterized by other, perpendicular
space coordinates.
We also want to stress that the procedure of tilting
the atomic surfaces is only a first-order approximation
to illustrate the idea. In reality, we should
introduce a kind of devil's stair case
in order to account for far-away changes
of configurations. Furthermore, we should modify
the atomic surfaces in such a way that it does
preserve the symmetry. Duneau has shown
that in an admittedly wrong, polynomial
approach for icosahedral symmetry, the minimal
degree of the polynomials
that comply with this
condition is three.
The idea of modulating the atomic surfaces
finds its confirmation in numerical calculations
of the dynamics, where it has to be introduced
in order to relax
the initial system.
The idea of modulating the atomic surfaces is also
present in a work of Steurer,\cite{Steurer} who calls
it the IMS setting (as opposed to the QC setting).

It remains to explain how the diffuse
scattering resulting
from our alternative model could decrease
when the temperature
is raised, but there are many possibilities to do this.
We have already stated that the data might
show that there
is some softening of the elastic constants.
Widom's instability, with tile-flip phason
elasticity replaced
by a phason elasticity based on small atomic
  displacements,
would already to the job. But it is even
not necessary
to claim that the QC would not be stable.
A mere softening of the elastic constants would do.
And it is even not necessary to invoke a softening
of the elastic constants. It could just be
that the system acquires supplementary
possibilities to reduce the strain.
One possibility is e.g. that thermal
vacancies contribute
to the relaxation of  the observed strain fields.
It is quite plausible that their number becomes
significant
at the temperatures where the fluctuations observed
by the authors set in.  Also fast phason hopping
between two positions
could help in relaxing strain fields.
For the physical origin of the continuous displacement
fields many causes could be invoked, e.g.
chemical disorder,
a domain structure
as described in reference,\cite{JNC} etc...
It is in view of this profusion of alternative
possibilities,
the unspecific character of the data
and the internal contradictions mentioned above
that the high-profile claims of the authors are just
not justified, especially as they are apt
to profoundly bias the opinions
on very crucial issues.

We may finally note that
a linear relationship
$\tau \propto 1/q^{2}$ certainly  does not prove
that the data are produced by a diffusion mechanism.
The relevant parameter for the diffuse
scattering is $1/q^{2}$, such that
the first terms of the Taylor expansion
for any possible underlying physics will lead
to $\tau = C_{1} + C_{2} (1/q^{2})$.
The authors
have previously considered the relationship
$\tau = C_{1} + C_{2} (1/q^{2})$
and this resulted in a much better fit
of their data.
Such a relationship is totally unspecific.
In general, fits of the type
$\Gamma= 1/\tau = \Gamma_{0} + D q^{2}$
(where $\Gamma_{0} \neq 0$ is an indication
for confined motion)
are used to analyze data when we already
know that they are produced by a diffusion mechanism,
not to present the data as supplementary
evidence that a diffusion
mechanism would be at work.

But let us admit that the analysis of the authors
in terms of a diffusion constant is correct.
We want to point out how
the use that the authors make of Lubensky's
statement that ``phason modes are diffusive''
is then still wrong and dangerously misleading,
as it creates the false impression
that the data analysis would contain further
proof for their interpretations.
The present author has tried to understand
the origin of Lubensky's statements,
by tracing back the citations
to ever earlier references, and
ended up in
specialized literature about liquid crystals.
Just as with the problem of
the microscopic interpretation
of the hydrodynamical theory one
feels thus being confronted with a black box.
Perhaps, Lubensky's statement means that
the hydrodynamical mode corresponds
to a diffusive type of dynamics.
In any case it is clear that
the  concepts of dynamics and hydrodynamical
modes should not be confused.
In other words, the fact that both are qualified by
the adjective ``phason'', does not imply
that ``phason elasticity'' has a 1-1-correspondence
with ``phason dynamics''.
E.g. Huang scattering in standard crystals is
traditionally
described as a ``frozen phonon'', but it has nothing to do
with  phonon dynamics.
The kinetics of frozen phonons will lead to
(very narrow) quasielastic scattering, while
dynamical phonons correspond in general to
non-zero frequencies.
{\em The kinetics of such frozen phonons shall
be ``diffusive'' in many instances, although phonons are
not qualified as diffusive by Lubensky.}
That the kinetics of the diffuse scattering is
of the relaxational or even of
the ``diffusive'' type is thus totally unspecific.
This reveals an improper, verbalist way of
using citations and homonyms, to validate personal
interpretations of
the terms ''diffusive'' and ``phason''.

These would be the essentials of our objections if it were
not that the paper dessiminates a number
of confusions  that are apt to interfere
with our attempts of clarification.
These confusions are centered on two main themes.

(1) Coherent scattering signals collect
contributions from all
particles of the system that have a non-zero
coherent scattering amplitude.
It is thus a many-particle signal.
Despite all possible folk lore, this should
not be identified with
a ``collective'' signal or with a signal
of some correlations,
as the latter implies that the particles
are no longer
independent and would move in a concerted,
correlated fashion,
due to some coupling or interaction, as is
e.g. the case for phonons.
In fact the oppositions independent vs.
not independent,
coherent vs. incoherent, single-particle vs.
many particle
cannot be amalgamated.

(1a) Coherent scattering with a certain
structure can also
occur in a system wherein the dynamics of all
particles are
totally independent.
A clear example of this are the Monte
Carlo simulations of
Tang et al. The tile flips are here completely
random and independent, but they lead to a clear
coherent signal that is actually very similar in its
reciprocal-space properties
to the one observed by the authors.
Certainly, it can occur that
locally, a tile flip only becomes possible
after another one,
and one could build a chain of such possibilities
over a long distance. But many other tile
flips can disrupt
this chain, and such chains are certainly not
the main contribution to the coherent
scattering signal
reported by Tang et al. What is the solution of the
apparent paradox that the dynamics are independent,
but nevertheless lead to highly structured
diffuse intensity?
In fact, totally uncorrelated jumps
in this model will nevertheless give rise
to strongly structured coherent signals, but
this is due to the {\em constraints} of the
random tiling
model, rather than some correlations or
lack of independence
in the tile flips. We should thus not confuse
the concepts of constraints and correlations:
The jumps are totally uncorrelated within the
given set of constraints dictated by the
random tiling model.
To give an analogon: For two completely
independent walkers
in a city where all streets run only eighter
North-South or East-West,
like New York (without Broadway), we might
find a mysterious correlation
in that they are found to walk always only
in mutually perpendicular or parallel directions.
This is not a mysterious correlation between the
two independent walkers, but a constraint imposed by
the city map of New York.

(1b) Coherent scattering signals can be obtained
from the dynamics of a single particle, provided
it is a coherent scatterer. This is
independent by definition.
Conversely, many-particle systems can
give rise to incoherent
signals, even if the dynamics are
strongly correlated.
It just suffices that the particles
are incoherent scatterers.

(2) A second amalgamate that should be avoided
is the one between waves and Fourier components.
QCs do not have translational invariance in the
form of lattice periodicity. It follows that
the Bloch theorem does not apply and the only
Bloch wave for an eigenvalue problem on a QC
is the trivial
constant function.  E.g. for a phonon problem,
the eigenvalue $\omega= 0$ gives rise to an
eigenvector
that takes the same values on all sites of
the quasilattice.
(But note that even this function
does not lead to a periodic structure after
its restriction
to the quasicrystal lattice: Its Fourier
transform is the diffraction
pattern of the quasilattice).
For $\omega \neq 0$ a ${\mathbf{q}}$-value
that one can read from
the pseudo-dispersion curve  merely defines
a Fourier component
of the eigenvalue ${\mathbf{v}}_{\omega}$:

\begin{equation}
\forall {\mathbf{r}}_{j} \in QC:
{\mathbf{v}}_{\omega}({\mathbf{r}}_{j}) =
\int\,{\cal{C}}(\omega,{\mathbf{q}})\,
e^{\imath {\mathbf{q\cdot r}}_{j}}\,d{\mathbf{q}},
\end{equation}

\noindent with ${\cal{C}}\in {\mathbf{C}}$.
  Experimentally,
we measure $|{\cal{C}}|^{2}$. For a periodic latice,
a dispersion curve
is a convenient representation of the
eigenvalues $\omega$ and
their corresponding eigenvectors
${\mathbf{v}}_{\omega}$
defined by ${\mathbf{v}}_{\omega}({\mathbf{r}}_{j}) =
e^{\imath {\mathbf{q}}_{\omega}
{\mathbf{\cdot r}}_{j}}$,
  as knowing ${\mathbf{q}}_{\omega}$
  is sufficient to
define the whole Bloch wave. For a
QC this no longer true.
Moreover, not also reporting $|{\cal{C}}
(\omega,{\mathbf{q}})|$ consists
in a severe loss of information, even if
it is true that the phase
factor of ${\cal{C}}(\omega,{\mathbf{q}})$ is
not available anyway.
It follows that phonons in QCs are not
properly studied by making
constant-${\mathbf{q}}$ scans and reporting
the corresponding
central value $\omega({\mathbf{q}})$ on a
  pseudo-dispersion curve.
One should rather make constant-$\omega$ scans and
report $|{\cal{C}}({\mathbf{q}},\omega)|$
for all ${\mathbf{q}}$-values that lead to
the same value of $\omega$,
such that a crude attempt to reconstruct
${\mathbf{v}}_{\omega}({\mathbf{r}}_{j})$
can be made (allowing for the fact that
without modeling we can never
have access to
the information about the phases of
${\cal{C}}({\mathbf{q}},\omega)$).
The quantity $e^{\imath {\mathbf{q\cdot r}}}$
for just one of these
${\mathbf{q}}$-values that define the
Fourier decomposition
of ${\mathbf{v}}_{\omega}({\mathbf{r}})$
has no physical meaning.
Analogously, in the speckle pattern, for
a given relaxation time
$\tau$, it is ${\mathbf{v}}_{\tau}$ that has
a meaning and not
a single component
$e^{\imath {\mathbf{q}}_{\tau}{\mathbf{\cdot r}}}$
that is being picked.
The quantity
$e^{\imath {\mathbf{q}}_{\tau}{\mathbf{\cdot r}}}$
is thus not a wave with any particular physical
meaning
but just a mathematical auxiliary quantity,
a Fourier component,
and Fourier components can be calculated for
mathematical
functions that have nothing of a wave,
e.g. a function that
is only non-zero on a given interval.
Moreover, such a Fourier component does not
correspond
to the ``simplified picture'' of a phason wave
introduced by the authors, as the
latter is {\em not} periodic.
And components with the same value
of ${\mathbf{q}}$ but
different values of ${\mathbf{Q}}$ may not have the
same relaxation time $\tau$.

In practice, the situation can be less bad.
The restriction to a quasilattic of a
periodic wave
that defines small (parallel-space) atomic
displacements,
will lead to many Fourier components with wave vectors
${\mathbf{Q}}+{\mathbf{q}}$. One may argue that
in the long-wavelength limit,  the
restriction of such a wave
may well lead to an intensity pattern
where ${\mathbf{q}}$  remains almost constant when
${\mathbf{Q}}$  runs through all the Bragg peaks.
The Fourier reconstruction of the initial wave
from the intensities while
keeping ${\mathbf{q}}$ rigorously  constant,
will then
not be too bad.
However, as in our reasoning we started from
a continuous wave of small atomic displacements,
we see once a again that one cannot
take it for granted that the reconstruction
of the intensity patterns observed by the
authors would lead to
a displacement field that consists uniquely of jumps.

One must make the distinction between a
  meaningful physical wave
and a mere Fourier component of an
instantaneous structural
configuration. The difference is that a
physical wave owes its pattern
to interactions between the atoms, while a
Fourier component can also
be defined for a system where all atoms are
mutually independent.
We can thus encounter three possibilities:
(a) a meaningful physical
wave as e.g. a phonon in a crystal;
(b) a Fourier component of a pattern
that is nevertheless produced by some
atomic interactions; and (c)
a Fourier component of a pattern that
comes about without interactions
because all particles are independent.
In the simulations
of Tang et al. we are in case (c),
while the diffuse scattering
observed by the authors we are
certainly in case (b) rather than
in case (a) as is being claimed.

In this respect, it might be misleading that
theoretical physicists also use
the term elasticity for situations where
there are no such
interactions between atoms. E.g. in the
simulations of
Tang, the tile flips are totally independent,
but still a ``phason elasticity'' can be
defined to refer to
an ``entropic restoring force''.
If this is not appreciated properly, it
can lead
to more confusion in terms of ``elastic wave''
pictures that do not apply.

(3) It is well known that a Katz-Kalugin
transition in the phason
dynamics is necessary to reach the full
random-tiling regime.
It is generally admitted that the few
claims that
this transition would have been observed
are ill-founded.

In conclusion, we have shown that the
claims of the authors
are not justified.
The claims are formulated in a terminology
that sounds very familiar but introduces
several confusions.

\section*{Discussion}

We reproduce here {\em verbatim} two referee
reports that were sent to us
by the editorial board of Physical Review and
used to thwart the publication
of our work.
We afterwards take their points one by one.
This is perhaps an unusual procedure, but it is
just too easy to
play on the appearances by writing
things that are very obviously wrong to an editor
who is not going to understand the issues anyway.
It is also just too easy on behalf of editors
to ``believe''
such very obviously biased reports uncritically.
\\

\noindent Report of the First Referee \\

The work of G. Coddens is devoted mainly to the analysis of
eventual mistakes in the the papers (cited as Ref. [1-2] in the
MS) explaining diffuse scattering in quasicrystals (QC) by
phason disorder.

 From the very beginning I'd like to
stress that Coddens'
arguments about the discrepancy in papers
[1-2] seem to me not
convincing enough. The arguments given in the
actual version of
the MS can permit simply to claim that there
can exist
mechamisms of diffuse scattering others
than phason disorder.

The mechanism of diffuse scattering generation
by small atomic
shifts proposed by G. Coddens represents in
its actual form an
abstract scheme which is not developed at all.
Maybe after a
detailed development of these raw ideas, namely
after a real
derivation of the expressions for scattering and
their real
comparison with the experimental data this work
will take a form
suitable for publication in PRB. But I doubt if
there exists any
serious scientific review which would accept the
MS in its
actual form. G. Coddens does not perform the
calculations to
obtain expressions for diffuse scattering but
proposes rather
general considerations. The reader is invited
to make the
conclusion (which seems to me rather doubtful)
that if one makes
all necessary calculations the mechanism
proposed by G. Coddens
would lead to the same expression as in
the classical model
(Ref. [2] in the MS).

Let me also briefly review the criticism of the
classical model
done by the author of the MS. First of all
this criticism is
based on the idea that in the traditional
model ``the number of
flipped tiles, and correspondingly the
diffuse scattering
intensity, should increase with temperature,
while the
experimental data exhibit precisely the
opposite temperature
dependence''. The situation is not exactly
like this. In the
framework of the traditional model, the
diffuse scattering
intensity should not necessarily increase
with temperature. One
should simply define this framework more
carefully:  At high
temperatures the overdamped phason waves
(elementary
non-localized excitations or phason
fluctuations with a
polarization in the perpendicular
space and a wave vector lying
in the parallel space) can be in the
thermodynamical equilibrium
with the system. The average of each
elementary wave of this
type is $1/2K_BT$. Each wave gives
the contribution to the
diffuse scattering amplitude
proportional to the scalar product
of the wave amplitude and  the
perpendicular component of the
corresponding Bragg reflection index.
Rather simple discussion
of above items can be found, for example,
in PRB 64, 144204
(2001)), by S.B. Rochal. The results
of Rochal are slightly
different from those obtained for the
first time in pioneering
work [2] as a generalization of Huang
effect for QC's.
Nevertheless, the difference between
both works is not essential
since the phonon elastic constants
in icosahedral QC's are much
greater than the phason ones (a big
number of publications prove
this fact). Due to this fact, the
amplitudes of phason
fluctuations (and their contribution
to the diffuse scattering)
are much greater than the amplitudes of
acoustic phonon
fluctuations. Then neglecting the phonon
contribution one easily
obtains the well known expression which
is used since at least
ten years by many authors (including
the authors of [1]) to fit
the diffuse scattering spectra. It's
difficult to suppose that
G.  Coddens is unaware of this formula.

According to this expression the
diffuse scattering intensity is
surely proportional to the temperature,
but it is also inversely
proportional to the phason elastic
constants. Thus, to explain
``precisely the opposite temperature
dependence''it is
sufficient to suppose that the
phason constants $K_1$ and $K_2$
increase with temperature faster
than the first degree of
temperature. The answer to the
question why this growth is fast
was given by Widom (Ref. [6] in the
MS) and is cited by the
author in the MS: ``In Widom's paper
it is stated that on
lowering the temperature the QC moves
to ... an elastic
instability''.

I'd like to add more comments (which
constitute a not exhaustive
list) on the MS:

1) The author tries to show that the
phason waves represent a
useless notion. Maybe it is the case
of the irrelevant phason
waves with a big amplitude considered
by the author. However,
the traditional model deals with the
phason waves with a small
amplitude.

In the second paragraph page 12,
G. Coddens gives the arguments
in favor of unphysical properties
of these waves but he does not
take into account the fact that
doubling of the wave amplitude
lead to a doubling of the switched
atoms number and,
consequently, to a doubling of the
superstructure reflections
amplitudes. I see nothing unphysical
in the properties of these
waves with the amplitudes proportional
not to the atomic
displacements but to the number of
switched atomic positions.

In addition, phason waves are the
solutions of the dynamic
equations and this fact justifies
their use in the theory.

2) According to the author, atomic jumps
correlated in space or
a diffision correlated in space are
impossible. But in this case
the order-disorder phase transitions
in solids are also
impossible, which is in evident contradiction
with all basic
physical data. Making a comment on the
consideration presented
in the end of page 12, I'd like to note
that the free energy of
an order-disorder transformation begins
with the quadratic terms
(like the elastic phason energy in QC does)
though  ``an atomic
jump never explores the harmonic regime of
the potential''.

3) The MS has no any structure.
It seems to me that the only
factor stimulating this work was ``a denunciation'' of the
authors of [1]. 3/4 of the MS is devoted to this aim.

Finally, I think that in its present form
the paper is
unsuitable for publication. I propose to
G. Coddens to develop a
constructive part of his theory. But I'd
like to stress that the
acceptable version of the MS should contain
the expression for
the diffuse scattering in this model and its
justified
derivation. Otherwise, my opinion on this
work will be
negative. If the author really tries to
perform this essential
calculation and if his final expression
is different with
respect to the classical model, it would
be very interesting to
see its comparison with the experimental data.\\

\noindent Report of the Second Referee \\

The manuscript submitted by G. Coddens
does not meet any
criterion for a publication in Phys. Rev. B.

- The manuscript is written in a
very aggressive style.  It is a
long and badly written paper against
the random tiling model
rather than a critic of the paper by
Francoual et al., whose
results were published in the framework
of the continuum
hydrodynamic theory of quasicrystals.

I develop below some of the arguments given above.

(i) The first seven pages of the manuscript
are a lengthy
discussion on the interpretation of papers
published 10 years
ago. In short, Coddens accuses Francoual et al.
to have built
up an ad-hoc theory to interpret the temperature
variation of
the diffuse scattering (an ``alternative random
tiling model'',
``180 degrees turn in data interpretation'').
Unfortunately this
point of view is wrong since the theory used
by de Boissieu et
al. in these papers is mainly the one of the
continuum
hydrodynamic theory, and the different
temperature ``scenarii''
were available in published simulations or
theoretical papers
(some of them not even referred to by Coddens,
as for instance
the key simulation done by Dotera and
Steinhardt, PRL, 1991).

Of course the question of the microscopical
interpretation of
the hydrodynamic theory is still an open
question. However, this
continuum theory is firmly established and
has been published in
many theoretical papers. In short, the
temperature dependence of
the diffuse scattering is related to the
still going one
controversy on the stabilising mechanism
of quasicrystal i.e.
energy versus entropy; two models have been
proposed which
predict a quite different behavior of the
diffuse scattering as
a function of the temperature: Coddens
simply ignores these
theoretical papers and prefers to criticize
the probity of de
Boissieu et al. who would have ``made a
180 degrees turn in the
interpretation''.

The ``random tiling theory'' is even not
properly presented by
Coddens. Contrary to what he writes the
``random tiling'' model
hypothesis is fully explained in a long
  paper by Henley
published in 1992: it is clearly stated
that the quasicrystal is
only stable at high temperature because of
configurational
entropy terms and that a transition toward a
crystalline state
should be observed. An ``inverse''
Debye-Waller factor is also
predicted by Widom and Ishii in their
study of the elasticity
theory of quasicrystals. Moreover the
author does not seem to
have fully understood the meaning of
these papers when for
instance p.7 he writes ``Widom's paper
talks about an inverse
Debye-Waller .... Why has this to apply
for the diffuse
scattering and not for the Bragg peaks'':
this is the well known
some rule which implies that any diminution
of the Bragg peak
intensity (by the Debye-Waller) results in
an increase of the
diffuse scattering.

(ii) The next two pages are dedicated to an
analogy with an
order-disorder transition. Whereas it might
be interesting to
look for order-disorder transition, especially
in metallic
alloys, the arguments given by Coddens have
little to do with
the complexity of what is encountered in
quasicrystals. Again,
there are false assertions: Coddens states
(p. 8 bottom) that a
phason mode with wavelength lambda;, should
show up at a
position $Q=2\pi/\lambda$; rather than at
$Q=Q_{B}+q$, where $q$ is the
wave vector of the phason mode and $Q_{B}$
the corresponding Bragg
peak position. Again this is wrong, and shows
a deep
misunderstanding of what is a phason mode.
All theoretical
papers dealing with phason modes clearly demonstrate,
that
equilibrium phason fluctuations are to be observed
at $Q=Q_{B}+q$, in
complete analogy with thermal phonon fluctuations
(references
12, in the PRL), and should display a continuous
distribution of
q wavevectors.

(iii) The remaining part of the manuscript
comes, again and
again, with repetitive arguments on the
microscopical
interpretation of the continuous hydrodynamic
theory. A few
pages are dedicated to an alternative
interpretation.
Unfortunately, Coddens failed to give any
credible arguments in
term of, for instance, a computation of
the hydrodynamic matrix,
or a simulation that would reproduce the
observed diffuse
scattering. Aware of this serious problem
Coddens writes: ``while
such a final, unambiguous solution is far
beyond our
possibilities we will nevertheless sketch
a few arguments to
convince the reader that an alternative
interpretation is not at
all impossible''. It is probably not
necessary to call on Karl
Poper for assistance to show that this
approach has little to do
with a scientific discussion. Moreover,
as for the other parts,
many mistakes can be outlined:

- First there is a poor understanding of
the hydrodynamic theory
by Coddens, a point that he recognizes
himself since he writes
`The present author has tried to
understand the origin of
Lubensky's statement.... And ended up in
specialized literature
about liquid crystals'`'. It is certainly
useful to recall that
the hydrodynamic theory has for basis
the famous theory of
``broken symmetry'', which by symmetry
and order parameter
arguments and analysis allows one to
predict the long wavelength
excitations that exist in a system. There
are many references
and even text book such the one of
Chaikin and Lubensky, in
which the theory is explained. Of course
this theory has been
applied in the field of liquid crystal,
but also in many other
condensed matter fields. We can quote for
instance: sound modes
in condensed matter, spin waves in
ferromagnets, second sound in
superfuid Helium4, supra-conductors....
One of the core
arguments of Coddens is a criticism of
the ``statement'' of
Lubensky (p.17 of the manuscript) that
phason modes are
diffusive. Unfortunately this is not a
statement but a rigorous
deduction from the broken symmetry and
mode counting analysis in
aperiodic crystal! A quite different story!
This as such is
enough to discredit the entire paper of
Coddens.

- The analysis of modes in quasicrystal
is wrong (p. 13 and
following). Coddens affirm (without any proof)
that a mode with
a wavevector in parallel, physical space,
and a polarization in
the perpendicular should be discarded.
This point has been
discussed in details by Radulescu and
Janssen in two papers:
indeed such a mode has to be considered,
both from an approach
in term of the hydrodynamic theory and
from simulations. Again,
instead of going either in some detailed
simulations, or
rigorous demonstration Coddens prefers
to stay on general
assessments in a non scientific posture,
without any critical
reading of the published literature.

- Coddens propose that small parallel
component to the atomic
surfaces might be an alternative and
that this might be related
to phason elasticity. That some parallel
components are
necessary to fully explain the structure
is well known both from
the experimental (note quoted by Coddens)
and simulation side.
However this parameter can not couple to
the phason elasticity
as claimed by Coddens. Indeed the very
essence of the ``phason
elasticity'' comes from the invariance
of the free energy of the
system with respect to a rigid translation
of the cut in the
perpendicular direction: this is this
degeneracy with respect to
this degree of freedom that leads to
phason modes. As shown by
Ishii for instance, this implies that
phason modes have a
polarization along the perpendicular
direction. Thus the small
parallel shifts are not linked with
the phason elasticity; they
are just a necessary relaxation to
minimize the energy of the
system by minimizing the energy of
the different local
environment.

(iv) The conclusion of the manuscript is
symptomatic of the only
objective of Coddens.

The paper by Coddens should be rejected.

\section*{Is it true?}

There are several points that illustrate
the lack of objectivity
of these referee reports. In fact,
a referee report should also point out
the parts that are correct. One such
point which is without any
appeal (see below) is that the diffuse
scattering that de Boissieu observes
cannot possibly correspond to the signal
of tile flip disorder as he claims,
because that is in contradiction with
model-independent neutron scattering data.
It is a point of major importance.
A long discussion of the paper is devoted
to this point, and all
referees pass it under the greatest silence.

Another illustration of the lack of
objectivity
are attacks on issues, to which the
answer is already clearly
given and discussed in the paper.
This way they present a very long
list of virtual objections  against my paper,
although they know that the answers to them
are already contained in the paper.
The sheer length of
this very efficiently creates the (false)
impression that there  is
an enormous amount of problems with my paper,
and that is of course extremely harmful
and detrimental to me.

Let me first make some introductory remarks.
The referee reports contain accusations
that I would question (1) Lubensky's
hydrodynamical theory,
(2) the theory of Jaric and Nelson,
(3) the random tiling theory based on
various papers.
That is total misrepresententation of the
issues and my paper.

There are two main issues in my paper.

(1) A theory might be very elegant and correct.
In physics the idea is that it
always must be
validated by experimental data.
That applies also to the random tiling model
in quasicrystals. Contrary to what one
would like to suggest, I have no pre-established
judgement about the validity of that
theory, and I totally
subscribe  the  results
of Lubensky and of Jaric and Nelson, on
which the random tiling theory is built,
because they are universally
valid, independently of the validity of
the random tiling model.

What I do have big trouble with is that
in case of the random tiling model,
the theoreticians take the work of the
experimentalist (A) as a proof for the
validity of the model (B), hence
$A \Rightarrow B$, while simultaneously
the experimentalist
uses the starting ansatz
that the random tiling model is correct
to interpret his data, hence $B \Rightarrow A$.
That is a vicious circle, and the reader
who tries to make up his mind is kept
running from Pontius to Pilatus.
It is the task of the experimentalist to
prove $A \Rightarrow B$,
but the motivation of the papers has been
largely dominated by $B \Rightarrow A$.
This way, it has been claimed time and again
that the random tiling model
would have been proved by the experimental data,
while these experimental
data all are almost unspecific. It is
totally
premature to make such high-profile
claims on the basis of such data.
Once again, this does not mean that I
claim that the random tiling model is wrong.
I am just stating that it has not been
proved, and that the statements
that one encounters, that it would have
been proved are falsehoods.

Physics is an inductive science, and it
is well known that
induction cannot logically justified in
a watertight fashion. Logically
spoken, induction is wrong. Hence, what
one does to improve
the credibility of an induction is in
general discarding as many
alternatives as possible.
An interpretation must therefore often
go with a lengthy discussion
of alternative possibilities,
trying to rule them out.
There is very little of that in the
claims $A \Rightarrow B$, and when
some of such subsidiary issues are debated,
in general they are concerned with
harmless, selected issues that
are not crucial. While, following Popper,
all issues should be examined.
One single discrepancy, one single counter
example is enough to condemn a
theory, how sexy it might be.
My viewpoint is not that the random tiling
theory would be wrong.
It is rather that it is still badly in
lack of proof.

(2) de Boissieu has superposed on (a)
his claims about the random tiling
theory the tenet (b) that the diffuse
scattering he observes is the
diffraction signal of disorder produced
by tile flips.
For me, the issues (a) and (b) are totally
disconnected, although
de Boissieu presents them as a single one.
Indeed, the quasi-elastic neutron
scattering data show that
the number of tile flips increases with
rising temperature,
while the diffuse scattering intensity
decreases with rising temperature.
The quasi-elastic data and their
temperature behaviour are
{completely model-independent information}.
They show that (b) is wrong.
Referee (2) tries to discredit me
by suggesting that this implies
that also (a) is wrong, and that the main contents of my paper
would consist in challenging
long-established theories, which is a total
mispresentation of the issues. There
is a whole lot of such methodology
of attacking my work on mispresentations of the issues I raise
in the report of referee (2),
as we will see. We will come back of
the origin of (b) below. \\

\noindent REFEREE 1.\\

(1) Referee 1 states ``The mechanism of diffuse
scattering generation by small atomic
shifts proposed by G. Coddens represents in its
actual form an
abstract scheme which is not developed at all.''
This remark can be classified under issue (2) above.
It can be rejected on several grounds

\noindent $\Rightarrow$ I have stated that
my quasielastic data
are model-independent information
that shows that de Boisieu is wrong on issue (2).
I could have stopped here. I have no logical
obligation to add anything further.
The argument is clean and without any appeal.

\noindent $\Rightarrow$ de Boissieu also
flagrantly contradicts himself when
he admits that the microscopic origin of
phason elasticity is not known,
while with (b) he stipulates exactly the
contrary, viz. that he
perfectly does know what this origin is,
specifying that the diffuse scattering
corresponds to tile flip disorder.

\noindent $\Rightarrow$ I tried to provide some
insight where de Boissieu
took an unjustified leap
in taking it for granted that the phason waves
from the theory would correspond
to tile flips (b), by giving a counter-example of
suggesting an alternative possibility.
According to this path of thought, de Boissieu
has to prove that
this alternative possibility is wrong in order
to prove that he can maintain his claim
that phason waves are tile flips, because it would be
the uniquely possible interpretation. It is
de Boissieu's task
to discard this possibility by a
calculation (actori incumbit probatio)
because he made the claim about the tile flips.
The query of referee (1) that I should make the simulation
is in this respect a reversal of the charge of proof.
It tries to impose on me a task that would
keep me busy for a long time, and that is just not fair.
It is remarkable that the same task has not been forwarded
to de Boissieu (see below), because what he claims
without proof by identifying (a) and (b)
is equally an abstract scheme that is not developed.

\noindent $\Rightarrow$ Nonetheless,
I have very good reasons to believe
that my alternative possibility is viable.
In fact, I proposed a QC model
wherein the atomic surfaces are tilted rather
than perpendicular to the cut.
That leads to diffraction patterns with the
very same Bragg peak positions
in both models, only the intensities are
different.
This is the only difference between the two
models. They are both quasicrystals.

Now considerations about the precise outlook
of the atomic surfaces
do not enter into the theories (1)-(3). All
one has to take care off,
is to make sure that the detailed nature of
the atomic surface
obeys to symmetry constraints. It is in this
respect that I cited
a private communication by Duneau. As the
issue if the atomic surfaces
are tilted (in the Fibonacci chain) or
perpendicular to physical space,
does not enter into the considerations,
the same theoretical machinery can be unleashed
on both models.
Hence, if there exists a detailed
calculation that shows that de Boissieu's
phason waves (in the way he defines them) lead
to the diffuse scattering
with the properties he observes, then by a 1-1
mapping of the arguments, the very same
calculation will show mutatis mutandis
that my model leads to diffuse scattering of the
same type. This argument is somewhat
similar to the one of Poincar\'e when he showed
that the parallels postulate in geometry
is independent, by making a 1-1-mapping between
the axioms of
non-Euclidean geometry
and Euclidean geometry. If de Boissieu is right,
I am right as well.

\noindent $\Rightarrow$ As we will see below,
by adding claim (b)
de Boissieu is taking an element
from an old version of the random tiling model
that is incompatible with
the version that is presently accepted on certain
crucial temperature
dependence issues.

In conclusion I would appreciate if referee (1) makes
himself the calculation he tries
to impose on me.  There is a very important reason to do that.
In fact, if one wants to maintain that issues (b) and
(a) cannot possibly
be disconnected, then the model-independent quasielastic neutron
scattering data show that the random-tiling model is wrong.
Rather than attacking the random tiling theory, as I
am being accused of,
I am trying to propose
a scheme that perhaps could
{\em save} the random tiling model from the
verdict of these data.
I could have left out this model, and it
would have been
much worse for the defenders of the random
tiling
model to figure out
a reply,
because in my present understanding it is their
firm believe that
(b) must be part of this model.
Perhaps, this may show that what I am interested
in are not personal
issues, as the accusations go, but in the truth.

(2) Referee 1 attacks me on the fact that
``the traditional model, the diffuse scattering
intensity should not necessarily increase with temperature''.
This is a completely false presentation of the issues,
and it contains the tacitly implied hint that I am so dumb
that I would not even understand this trivial point.
The whole paragraph that contains this statement
and the next one are useless.
They are very long, creating the impression
that there is a whole lot
wrong with my paper, but it is not about my paper.
I already pointed out why above: I am relying
on model-independent data
and I am not attacking any theory.
Moreover, the point is clearly stated in my paper,
but the referee passes it under silence.
My paper does not adress the issues the referee tries to
put on my back. I am only forced to cite them
because of what de Boissieu writes.

\noindent $\Rightarrow$ First of all that
statement comes from de
Boissieu himself, as can be seen
from his ISIS report.

\noindent $\Rightarrow$ Secondly, before
Widom's paper (Theory 2)
was published, the random tiling theory
already existed. Widom's paper introduced new features,
that were not accounted
for in the pristine theory (Theory 1). Experimental
data related to these new features
would have not been accounted for by the pristine
theory.
In fact, while Theory 1 foresees that the diffuse
scattering
intensity should increase when the Temperature is
increased [x],
Theory 2 states that the intensity can both
increase [x] or decrease [y]
when the temperature is increased.
In this respect Theory 1 and Theory 2
contradict each other on issue [y], because
Theory 1 denies the possibility [y]. It is exactly to this point
that de Boissieu's statement refers, because
his data clearly indicate
that we are in the presence of [y].
Now referee 1 misrepresents the whole discussion, by
pointing out that the opposition between [x] and [y]
is meaningless within Theory 2, while it is
perfectly meaningful
between Theory 1 and Theory 2.

\noindent $\Rightarrow$ We may note that it is
{\em within the motivation} of
Theory 1 that
the result of Tang has been derived. I
emphasize on the motivation, because
Tang's result is model-independent. Tang has
shown that tile flip
disorder (phason1) leads to diffuse scattering
of a type that is comparable
to what can be deduced from a model with
continuous phasons (phason2),
following Jaric and Nelsson,
while the tile flip disorder (phason 1) is
something entirely different from
continuous phasons (phason 2).
Hence:\\
phason1 $\Rightarrow$ diffuse scattering (Tang)\\
phason2 $\Rightarrow$ diffuse scattering
(Jaric and Nelsson).\\
while  not(phason1 $\Leftrightarrow$ phason 2)
(else Tang's work would not have
been necessary, because it could have been derived
from the equivalence
and the work of Jaric' and Nelsson).

\noindent Due to the fact that

$\neg$ (phason1 $\Leftrightarrow$ phason2)

\noindent it is obvious that both

diffuse scattering $\Leftrightarrow$ phason1

diffuse scattering $\Leftrightarrow$ phason2

\noindent are logically wrong. Additional
information is required in
order to conclude what kind of ``phasons''
are signalled by the diffuse scattering.
Nevertheless, de Boissieu claims without
any proof or discussion
that the diffuse scattering corresponds
to phason1, which is his claim (b).
He even claims that phason1 and phason2
are equivalent, because
tile flip disorder would be waves of
correlated atomic jumps.

The worst is that all this is perfectly
present in the paper
despite all conclusions one might be
tempted to draw from the referee's presentation.

(3) Referee 1 states:``The author tries
to show that the phason waves represent a
useless notion. Maybe it is the case of
the irrelevant phason
waves with a big amplitude considered by
the author. However,
the traditional model deals with the phason
waves with a small
amplitude''.

Again this does not be discussed, beacuse it
is perfectly
present in the paper, when I point out that
on going from 6D to 3D
the wave character is lost.
Here referee 1 persists in denying an obvious
and trivial mathematical
fact, viz. that the mapping that translates the
6D wave to a 3D
atomic displacement field is not analytical.
The effect of an infinitesimal variation of the
6D wave
is not infinitesimal in 3D: it is either 0 or a
finite number (the minimal phason jump).

What I maintain is that phason waves are a
useless motion in the context of (b),
i.e. that phason waves are a meaningless concept
in (phason1), not in (phason2). Nevertheless,
the referee uses (phason2)
to attack me on statement that is within the
context of (phason1).
Again the assertion of the referee is not true,
and again I have discussed it in the paper.
And, as the reader can check, I have discussed
both small and large amplitudes.

When the amplitudes are small,
there will exist only a very tiny amount of
flipped tiles, separated by big distances.
That is not a physical wave:
In a real wave there is some displacement
amplitude in every lattice site of its extent.
One can then understand that the displacement
of a first atom,
induces a displacement of a second atom, etc...
And this way one can build a chain
of cause and effect over large distances.
That is not at all the case here:
the atoms in between remain exactly where
they were, there are large intermediate regions
without any flip at all, and
it is thus impossible to understand how
the flip in the first position could have
induced one in the remote second position,
as there is no neighbour-to-neighbour
transmission mechanism.
There can thus be no
force transmitted between such atoms that
are separated by a large distance,
with nothing in between.

The same displacement field could be obtained
by a totally different perp-space function
than the sine wave proposed.
Morover, the Fourier components in the
diffraction pattern
of such a displacement field will not
correspond to the wave initially postulated.
It is easy to draw of a small-amplitude
sine-wave where there are no
atomic flips over a whole period, and where
one has to go out much further
out in space to find the first next flip.
The doubling of the amplitude of the 6D wave,
will not
lead to an exact doubling of the intensities of
the Fourier spectrum,
but add new Fourier components, etc...
What the referee states:``but he does not
take into account the fact that doubling of
the wave amplitude
lead to a doubling of the switched atoms number and,
consequently, to a doubling of the
superstructure reflections
amplitudes'' is rigorously wrong, because for
an exact doubling
of the Fourier amplitudes, one has to double
the effect in direct space
in exactly the same positions. Adding switched atoms
  in new positions is not such a doubling. And
the doubling of the number of switched atoms
is not rigorous.

He also states: ``In addition, phason waves
are the solutions of the dynamic
equations and this fact justifies their use
in the theory''
This is a mere play with words, that consists in
confusing the reader between (phason1) and
(phason2). The microscopic interpretation of
the phason waves that are the solutions of the
dynamical equations (which yield (phason2))
is not clear, as de Boissieu states.
Therefore it is not clear
if these waves correspond to tile flips (phason1),
as the author takes for granted
by taking advantage of the fact that the word
``phason'' has been used
with several different meanings. That is
also clearly written in the paper,
and again the referee prefers to act as
though he has not seen it.

The phason waves of small amplitude the
referee refers to
are in the infinitesimal-amplitude limit,
while there is even
nothing that permits to think that in de
Boissieu's
tile flip disorder scenario
the amplitudes of the  Fourier components
of the function
that describes the  deformation of the cut
remain small.
There is no proof that the amplitudes in de
Boissieu's tile flip scenario
are not large. And his data even do not
correspond to the tile flip
scenario. Finally, the problem is not if
the displacement field
proposed by de Boissieu is physical, but
that it is not a wave, as
he claims. His displacement field does not
correspond
to the wave theory he tries to invoke.

(4) referee 1 states: `` But in this case
the order-disorder phase transitions in solids are also
impossible, which is in evident contradiction
with all basic
physical data.''

This is reasoning against obvious facts.
What shall we think about a claim without
any proof that in a metallic alloy, atomic
jumps are correlated
over large distances, e.g. 1000 or 10000
Angstroms.
And  what should we  think about the reply
that my objection  would be mere speculation,
when I am pointing out that
postulating such
{\underline {\bf {\em coherent diffusion}}}
is a strange, exceptional claim,
that, to use an euphenism, requires
discussion and proof.

There is no prove for the existence of a
phase transition in QCs.
Let alone that it would be an order-disorder
one.
This is typical of a whole attitude
of just layering one ad-hoc assumption on
top of another one.
There is no evidence for a phase transition
in QCs, as is crucial for the random
tiling model? Surely, it just is not
observed due to the slowing
down of the kinetics. There is no reason
for believing
that phason jumps are correlated over long
distances in QCs?
Surely there must be a order-disorder
transition....

Let us ironize a little bit
about the claim about
long distance correlations.
Let us consider two points
A and B separated by a large
distance of 1000 Angstroms.
In point A we have a first jump.
Due to the correlation invoked,
then one of its neighbours will jump.
Due to the correlation invoked,
then one of the neighbours of the latter
one will jump, etc...
This way we build a whole domino game of
correlated jumps
diffusing from A to B. The whole chain of
causes and effects
arrives at B after about 100 seconds,
inducing a jump at B.
The problem is that the neutron and M\"ossbauer
data show that at
the temperature of 650°C
of his experiment, the atomic jumps
take place on the time scale between picoseconds
and nanoseconds.
Hence, during the 100 seconds time lapse de
Boissieu is talking about,
the atom in B will have jumped a $10^{11}$ to $10^{12}$
times that cannot be attributed
to a correlation that would link a jump in B to a
jump in A, by diffusing from
B to A.

And as I pointed out under (3), when the waves
de Boissieu stipulates
have very small amplitudes, they
have nothing to do with such a neighbour-to-neighbour
transmission mechanism,
because they contain large regions where there
are no flips at all.

(5) referee 1 states: ``I'd like to note that the
free energy of
an order-disorder transformation begins with
the quadratic terms
(like the elastic phason energy in QC does)
though  ``an atomic
jump never explores the harmonic regime of
the potential'' ''.

This is again reasoning contrary to obvious facts.
The very nature of an atomic jump implies that it jumps
in a double-well. Such a double well can never be
described by a quadratic function.
The atom may explore harmonic forces when it is
at the bottom of
one of the double wells, but not on the saddle point on
its way to the
other well. So, whatever his mental construction tries to
prove, it runs contrary to very
obvious facts.

(6) Referee 1 states: ``It seems to me that the only
factor stimulating this work was ``a denunciation '' of the
authors of [1]. 3/4 of the MS is devoted to this aim''.

This is again a misrepresentation of the issues.
He could also have  written ``a disatisfaction
with the arguments of [1]'',
because all that counts is the validity of my
scientific arguments,
and the personal motivations he wants to accredit me
with should not have any incidence
on the judgement of my paper.

(7) referee 1 concludes: ``I propose to G. Coddens to develop a
constructive part of his theory. But I'd like to stress that the
acceptable version of the MS should contain the expression for
the diffuse scattering in this model and its justified
derivation. Otherwise, my opinion on this work will be
negative. If the author really tries to perform this essential
calculation and if his final expression is different with
respect to the classical model, it would be very interesting to
see its comparison with the experimental data.''

This would just cripple my paper and conveniently hide
a few very disturbing facts from the eyes of the community.
What we can keep from it is that we can  see that
the referee admits that
he has no arguments to claim it would be wrong.\\

\noindent REFEREE 2 is de Boissieu.\\

In this report there is a systematics of
misrepresenting
the issues I raise. Each time my issues
are reformulated and denaturated
in the form of an issue that is so very
obviously wrong, that you could not
possibly win, if you yielded to the temptation
to get sidetracked on this false issue.
Referee (2) advances a whole collection of such
false issues and shows
each time with pathos
how ridiculous and incompetent they are.
That creates a very bad impression about
my paper (not to say me), but it all has
nothing to do with the real contents of my paper.
The bottom line of it is that we are confronted
with one elusive move
after another one, and that it
appears impossible to pin him down on a discussion
of the crucial real issues.
This clearly shows that de Boissieu has no answer
to the issues I am raising.
The many {\em ad hominem} statements also attempt to
elude
a discussion of these issues on a scientific
level,
by having the paper rejected on the basis of
the non-scientific issues.

(1) He states: ``It is a
long and badly written paper against the
random tiling model
rather than a critic of the paper by Francoual
et al., whose
results were published in the framework of
the continuum
hydrodynamic theory of quasicrystals.''

I have already stated that this is a
misrepresentations of the issues above.
I am not attacking the random tiling model.
The words ``in the framework'' confirm, what I
pointed out above,
viz. that de Boissieu uses B $\Rightarrow$ A.
In fact de Boissieu works ``in
the framework of the random tiling model''.

(2) In the ISIS report de Boissieu
admits that he would have expected the opposite
temperature behaviour
and that he had to shift to a more elaborated
random-tiling model.

Afterwards this temperature behaviour has
been used
to claim that it proved the random tiling
model.
As the same could have been claimed with
the opposite behaviour,
this has no persuasive value at all. de
Boissieu should
have tried to prove that this temperature
behaviour
excludes all possible alternative models,
which he did not.
In stead of that he tacitly introduced
a ``tertium non datur''
and went on claiming. As I explain in the paper
most of the results he uses do not discriminate
between
the random tiling and alternative models at all,
as they rely on Lubensky, and on Jaric and
Nelsson,
while these papers have a much larger scope
of generality
and remain perfectly valid in the context of
all possible
other models.

It has been claimed that the temperature
behaviour would contradict
the energetic stabilization scenario, which is really
an oversimplification of the issues, by
presenting it as though it
could not be that the energetic stabilization
scenario
could be right, altough it could be  something
totally different than the
stability issue which
is producing the effects observed in the data.
That the defenders of the energetic scenario do not reply,
is not because they could not possibly think
of a scenario.
It is because that there are so many scenarios,
that
it would be ridiculous to pick one and propose
the one picked as the one.
There is simply not enough information to
allow picking one.
The entropic stabilisation scenario has
difficulties in its own
right because it stipulates a periodic ground state,
which is not observed, and then introduces the
ad hoc assumption
that the ground state is not reached because
the kinetics slow down
too fast.
All this shows is that the whole issue of energetic
against entropic scenario
is far more complex than de Boissieu's tries to suggest.
He tries to impose that issue by brute force
on his data suggesting that
  the mere temperature
dependence of the diffuse scattering would be
able to settle the issue.
It is not because some data can be put in a
given perspective
(e.g. a stability issue) in one theory, that
it must be put
in the same persective
in a competing theory. They could address a
totally different
issue in the competing theory.

(3) It is then stated: ``Coddens simply ignores
these
theoretical papers and prefers to criticize
the probity of de
Boissieu et al. who would have ``made a 180
degrees turn in the
interpretation''. ''

Once more this has nothing to do with my
paper, as my point is model-independent.

(4) It is stated: ``Contrary to what he writes
the ``random tiling'' model
hypothesis is fully explained in a long paper by Henley
published in 1992''.

The phrasing ``Contrary to what he writes'' is
just a falsehood,
and tries to put words into my mouth that I never said.

(5) ``Moreover the author does not seem to
have fully understood the meaning of these papers
when for
instance p.7 he writes ``Widom's paper talks
about an inverse
Debye-Waller .... Why has this to apply for the diffuse
scattering and not for the Bragg peaks'':
this is the well known
some rule which implies that any diminution of
the Bragg peak
intensity (by the Debye-Waller) results in an
increase of the
diffuse scattering.''

This is again a misrepresentation.

That the diminution of the Bragg peak in the
data goes with the increase
of the diffuse scattering is clear in the
experimental data, and as such this is model
independent. Contrary to what de Boissieu claims,
I understand this perfectly,
and it is even written in my paper, when
I state that the intensity
is transferred from the diffuse scattering
to the Bragg peak.

A true Debye-Waller diminishes all the elastic intensity
at the profit of the inelastic intensity. There is
a sum rule between the
elastic and the inelastic intensity.
My point was that the Bragg peaks and the diffuse
scattering intensities are {\em both elastic} intensities,
such that they should be both affected the same way.

I have found the solution to that riddle myself
in the mean time.
What Widom calls the Debye-Waller factor in his paper
is not the true Debye-Waller factor, but a theoretical
auxiliary quantity that consists on integrating on only
the very long-wavelength modes (i.e. on only an
infinitesimal
domain of q-vectors). It thus excludes the whole
phonon density of states (as e.g. measured by Suck),
except
that infinitesimal part. Similarly, it excludes all
the phason
dynamics I have measured by TOF neutron-scattering,
except an infinitesimal part.
Within this long-wavelenth approximation Widom
then calculates
his auxiliary quantity that indeed applies to
the transfer of
intensity between the Bragg peak which is elastic and the
diffuse scattering which becomes inelastic in
this approach.
But it follows that the true Debye-Waller factor,
obtained by
including the rest of the full q-range, could
again invert the tendencies described by Widom, i.e.
some Bragg peaks could show the inverse effect, while
others could show the normal effect.

(6) ``The next two pages are dedicated to an
analogy with an
order-disorder transition. Whereas... should
display a continuous distribution of
q wavevectors.''

Again this is a total misrepresentation of
what I am saying.
de Boissieu is putting here words into my mouth that
I never said.
  I have not written anything like that at all.
  As can be seen on the place
he refers too, I was talking about spin waves.
The context is totally different from what he
tries to make us believe.
I am adressing the point that according
to the random tiling
model there should be a low-temperature periodic phase,
and that this phase has never been observed.
The primordial claim of the random tiling, the
low-T periodic phase has not been observed!
The ad hoc assumption that has been used to talk
this serious objection away
is that the low temperature phase is not
reached due
to a quenching of the kinetics.

I inspect therefore if some observations could
nevertheless be used
to claim that a phase transition has indeed
been observed.
I see two possibilities: (a) a periodic phase
like the rhombohedral
microcrystalline approximant in AlCuFe; (b)
the phase with the satellites
observed by Ishimasa.
First of all these are exceptions rather
than the rule. For a slightly
different concentration the phase transition
is missed.
Secondly, the rhombohedral phase has a large
unit cell.
Such large unit cells are hardly an improvement
over the QC
in the question on how they can possibly exist.
Third, in Ishimasa's phase we get sattelites,
which means that
we go to a modulated QC phase, which is in
fact even more complicated
than a periodic phase, and cannot serve as
the periodic ground state, we are looking for.
Finally, and this were de Boissieu totally
misrepresents the issues,
if the phason waves are the unique mechanism
that drive
the phase transition to the periodic ground
state, they should build up progressively intensity
at the future Bragg peak positions of the
periodic ground state, and no such
tendencies can be observed in the data.
The Q-values of these Bragg peaks have
a physical meaning in the periodic low-T system.
And the q values of intensity at positions
q+Q, where Q is such a future Bragg peak,
have also a physical meaning,
with respect to the low-temperature phase.
For sure, in the periodic phase,
there are no phasons at all.
When a phase transition takes place, we are
entitled to relate
at a certain stage of progress in the transition
the Q-values to the frame
of the periodic phase rather than to the frame of
the quasiperiodic phase,
and the interpretation of intensity must change
from phason to non-phason.
de Boissieu tries by brute force to discredit
my remarks by
imposing the quasiperiodic reference frame,
and imposing the framework
of the random-tiling theory on top of that,
while my remarks are
in a different context, and not stupid as
he would like us to believe.
I have never stated that the distribution would
not be continuous,
as he tries to make us believe. I have only stated that it
should contain local maxima, which are not observed.

I think my motivations were perfectly spelled
out in the paper,
but once agin they have just been totally
misrepresented.

(7) referee (2) states:

``Aware of this serious problem Coddens
writes: ``while
such a final, unambiguous solution is far beyond our
possibilities we will nevertheless sketch a
few arguments to
convince the reader that an alternative
interpretation is not at
all impossible'. It is probably not necessary to
call on Karl
Poper for assistance to show that this approach
has little to do
with a scientific discussion.'' ''

As I pointed out above, I do not have to make
this calculation
to prove that his assertion (b) is wrong.
He should have made this calculation to prove
that assertion (b) is right
(which it is not anyway) by eliminating the
alternative.

(8) There is the passage:
`` First there is a poor understanding of the
hydrodynamic theory
by Coddens, ... This as such is
enough to discredit the entire paper of Coddens.''

This is again a misrepresentation of the issues.
Lubensky's presentation does not permit us to spot
the rigorous definitions that would allow
us to invalidate the
use that is made of the statement that
phasons are diffusive.
I am not at all questioning the symmetry
arguments of Lubensky as de Boissieu
wants us to believe. I am just
pointing out that the statement that the
phason modes
are {\em diffusive} has been misused by de
Boissieu, by using a play of words.
It must be that
two different meanings of the word ``{\em diffusive}''
are identified,
because diffuse Huang scattering scattering
based on phonon rather
than on phason elasticity, would also manifest
itself as ``{\em diffusive}''
(in the sense he uses it)
in the type of experiment de Boissieu reports, while
phonons are not diffusive
in the sense of Lubensky.
de Boissieu is going to extremely exotic situations
(time scales of minutes) to claim that identification,
while there are many other time scales where
Lubensky's theory should be right as well.

(9) ```The analysis of modes in quasicrystal is wrong ...
without any critical
reading of the published literature.'' is again a
misrepresentation.
The only thing I said was that once we project the physics
onto 3D space (which one always has to do in the end),
there must remain some polarization in the physical space,
else the wave is physically irrelevant.
de Boissieu uses the fact that I swap between a 6D mathematical
and a projected 3D
physical description to present my language in 3D as
though it would be
stupid language
in 6D. The two contexts have to be clearly separated,
which I do in the paper,
but he mixes them up to discredit me.

(10) ``Coddens propose that small parallel component ...
energy of the different local
environment.''

I have already given the explanation of my model
above, showing that it is in 1-1-correspondence with his claims.

Finally, I would like to point out the reversal of r\^oles
and denial of rights of recourse
installed by the accusation that I would
  raise personal issues.
For more than 12 years Janot, Dubois and de Boissieu
systematically questioned my competence with an incessant
series of far-fetched invalid criticisms on my work
on quasielastic neutron scattering
from quasicrystals (phason hopping):

\noindent (a) the quasielastic signal would not be due to atomic hopping
but to localized vibrations from clusters.

\noindent (b) the quasielastic signal would not be due to phason hopping
but due to preferential segragation of Cu into the grain boundaries.

\noindent (c) the quasielastic signals are not due to phason hopping
but to rotating molecules.

\noindent (d) the quasielastic signals would be minuscule and could only be
evidenced with the aid of a fit program.

\noindent (e) phason hopping is nothing special for quasicrystals.

\noindent (f) phason dynamics occurs also in periodic crystals.

\noindent (g) the quasielastic signals could be due to tunneling states.

\noindent (h) The important issue in phason dynamics is not
the hopping but long-wavelength phason fluctuations.

\noindent (i) The first experimental evidence for
phason hopping was obtained by Janot and
published in ILL reports.

We submitted
in 1994 a proposal to the ILL to measure phason dynamics.
In the selection committee Dubois  told
the other committee members that
the experiment had already been done
by Janot, and therefore should
not be awarded beam time. It was just not true.
Our proposal was then rejected without even being discussed.
When I objected, I was discredited,
just like now, with the argument that I tried
to  question the probity of this group and of the entire ILL.
A few months later Janot and de Boissieu did the
experiment in our place in ``test'' beam time on IN16,
but they melted their sample.
This attempt is e.g. mentioned in reference \cite{wrong2}
and the citation [18] therein.
Contrary to the self-acquitting claims in these papers,
they had searched for the very
same signal, in the same Q-range, in
the same very narrow energy-range,
within the same temperature range, on the  same
single-grain QC alloy, with the same type of instrument.
  They had these  claims
co-signed by two colleagues from my own lab,
who discovered {\em post factum} that
their request to remove the corresponding passages
had been ignored. To prove that it was
``different'' they introduced the wrong
alternative  interpretation they
  wanted to give
to the type of quasielastic signals we had measured,
replacing the scientific issues, and shifting
appreciations.

The undue confusion produced by (a)-(i) tends to ease
  the attempt to install this shift
through reference \cite{Fracoual}.
Each time I had to discover
these informal and damaging Comments
(with their {\em de facto} denial of my rights of reply)
as an accomplished fact in the published literature.
To try to undo the damage I was forced to write a Comment
with reversed  rights of reply.
Their replies contained then new invalid damaging
statements, to which I was, as always, denied any possibility
of recourse. I really do not see, why I should have to accept
being framed up with such methods, and moreover being
personally discredited
on the basis of totally biased representations of the context
and one-way readings of the rules,
each time I try to defend myself against this.


\begin{thebibliography}{99}


\bibitem{Fracoual} S.~Francoual, F.~Livet, M.~de Boissieu,
F.~Yakhou, F.~Bley, A.~L\'etoublon, R.~Caudron, and J.~Gastaldi,
Phys. Rev. Lett. {\bf 91}, 225501 (2003);
A.~L\'etoublon, F.~Yakhou, F.~Livet, F.~Bley,
M.~de Boissieu, L.~Mancini, R.~Caudron, C.~Vettier and J.~Gastaldi,
  Europhys. Lett. {\bf 54}, 753 (2001).

\bibitem{Jaric} M.V.~Jari\'c and D.R.~Nelsson, Phys. Rev. B {\bf 37},
4458 (1988).

\bibitem{Tang} L.H.~Tang, Phys. Rev. Lett. {\bf 64}, 2390 (1990).

\bibitem{Shaw} L.J.~Shaw, V.~Elser and C.L.~Henley,
Phys. Rev. B {\bf 43}, 3423 (1991).

\bibitem{ISIS} M.~de Boissieu, M.~Boudard,
and C.~Janot, in {\em ISIS 1994-1995}, (Rutherford Appleton Laboratory),
p. A23. The authors report here their data obtained on the instrument SXD
(July 1992).

\bibitem{Widom} M. Widom, Phil. Mag. Lett. {\bf 64}, 297 (1991).

\bibitem{Popper} K.R.~Popper in {\em The Logic of Scientific 
Discovery}, 9th edition,
(Hutchinson, London, 1977), p.40.

\bibitem{wrong1} M.~de~Boissieu, M.~Boudard, B.~Hennion,
  R.~Bellissent, S.~Kycia, A.~Goldman, C.~Janot, and
  M.~Audier;  Phys. Rev. Lett. {\bf 75}, 89 (1995).

\bibitem{IJMPB} G.~Coddens,  Int. J. Mod. Phys. B {\bf 7}, 1679 (1997).
This paper raised already many objections against
the tile-flip interpretation of the diffuse scattering,
but the authors ignored it and just repeated their claims.

\bibitem{Elser} V.~Elser, Phil. Mag. B {\bf73}, 641 (1996); C.L.~Henley,
oral presentation at the {\em NATO Advanced
Study Institute on Mathematics of
Long Range Aperiodic Order. The Fields Institute of Research
in Mathematical Sciences, Waterloo, Ontario,
Canada, August 21 - September 1, 1995}.

\bibitem{private} M. de Boissieu,
unpublished correspondence with the
editorial board
of Physical Review Letters.

\bibitem{remark} The time scales (i.e. the energy spectra) of these
intensities may change with temperature,
but this cannot be seen on the X-ray diffuse
scattering data, which have very course energy resolution.

\bibitem{Ishimasa} T.~Ishimasa, Y.~Fukano, and M.~Tsuchimori,
Phil. Mag. Lett. {\bf{58}}, 157 (1988)

\bibitem{speculation} M. de Boissieu, unpublished
correspondence with the editorial
board of Physical Review Letters.

\bibitem{Salje} A.M.~Bratkovsky, S.C.~Marais, V.~Heine and E.K.H. Salje,
J. Phys.: Cond. Matter {\bf 6}, 3679 (1994).

\bibitem{warning} However, the statement that the other elastic constants
do not intervene, should not be taken as absolute.
Else we risk to complete our modeling efforts
by trying to comply to an absolute constraint,
that might make our attempts prohibitive.
There could be small contributions from the other constants.

\bibitem{Henley} C.L.~Henley in {\em Quasicrystals, The State of the Art},
5World Scientific, Singapore, 1991), p. 429.

\bibitem{myPRL2} G.~Coddens, S.~Lyonnard, B.~Hennion, and Y.~Calvayrac;
Phys. Rev. Lett. {\bf 83}, 3226 (1999).

\bibitem{wrong2} M.~Boudard, M.~de~Boissieu, A.~L\'etoublon, B.~Hennion,
R.~Bellissent and C.~Janot,  Europhys. Lett. {\bf 33}, 199 (1996).

\bibitem{note}
A propagation along perpendicular
space of the type
$\sin({\mathbf{q}}_{\perp}{\mathbf{\cdot r}}_{\perp}-\omega t)$
does not really lead to physical motion,
except for the short time interval when
the sine wave will traverse the neighbourhood of the cut.
Most of the time it will
affect only atomic surfaces that are far away from the cut
and are thus irrelevant for the actual structure.
But this objection can be overcome by
noting that following the statements of Lubensky,\cite{Lubensky}
we have already given up on the notion of propagating wave.\cite{Lubensky}
The displacement field will then exhibit a characteristic
decay time $\tau$ to diffuse completely out of that neighbourhood of the cut
that leads to real atomic displacements.



\bibitem{JNC} G.~Coddens~G., and
R.~Bellissent, J. Non-Cryst. Solids
{\bf 153 \& 154}, 557 (1993).

\bibitem{Steurer} W.~Steurer, Z. Kristallogr. {\bf 215}, 323 (2000).
However, it is in general not rigorous
to introduce the notion that there exists
  a 1-1-mapping
between a quasicrystal and a periodic structure.

\bibitem{Lubensky} T.C.~Lubensky in{\em Introduction to Quasicrystals},
ed. by M.V.~Jari\'c, (Boston Academic Press), 1988.


\end{thebibliography}
\end{document}